\documentclass[12pt]{article}
\usepackage{graphicx}
\usepackage{cite}
\usepackage{color}
\newcommand{\bm}[1]{\mbox{\boldmath $#1$}}
\newcommand{\fnd}[2]{\frac{\textstyle #1}{\textstyle #2}}

\newcommand{\dissum}[2]{\displaystyle \sum_{#1}^{#2}}
\newcommand{\Real}[1]{\Re {\it e}\left(#1 \right)}
\newcommand{\Imag}[1]{\Im {\it m}\left(#1 \right)}

\begin{document}
\title{The complex relation between production and scattering amplitudes}
\author{
Eef van Beveren\\
{\normalsize\it Centro de F\'{\i}sica Te\'{o}rica,
Departamento de F\'{\i}sica, Universidade de Coimbra}\\
{\normalsize\it P-3004-516 Coimbra, Portugal}\\
{\small http://cft.fis.uc.pt/eef}\\ [.3cm]
\and
George Rupp\\
{\normalsize\it Centro de F\'{\i}sica das Interac\c{c}\~{o}es Fundamentais,
Instituto Superior T\'{e}cnico}\\
{\normalsize\it Universidade T\'{e}cnica de Lisboa, Edif\'{\i}cio Ci\^{e}ncia,
P-1049-001 Lisboa, Portugal}\\
{\small george@ist.utl.pt}\\ [.3cm]
{\small PACS number(s): 11.80.Gw, 11.55.Ds, 13.75.Lb, 12.39.Pn}
}

%11.80.Gw 	Multichannel scattering
%11.55.Ds 	Exact S matrices
%13.75.Lb 	Meson-meson interactions
%12.39.Pn 	Potential models

\maketitle

\begin{abstract}
The unitarity relation $\Imag{A}=T^{\ast}A$ is derived for a three-body
production amplitude $A$ that consists of a complex linear combination of
elements of the two-body scattering amplitude $T$. We conclude that the
unitarity relation does not impose a realness condition on the coefficients
in the expansion of $A$ in terms of $T$.
\end{abstract}

Under the spectator assumption, we deduced in Ref.~\cite{ARXIV07064119} that
the three-particle production amplitude $A$ consists of a complex linear
combination of elastic and inelastic matrix elements of the two-body
scattering amplitude $T$. Furthermore, in Ref.~\cite{JPG34p1789}
we showed that such a two-particle production amplitude can reasonably
describe experiment in an energy region where no additional resonances from
possible rescattering with the spectator particle exist. Moreover, no need to
treat a sizable fraction of the experimental signal as background was noticed.

The result of Ref.~\cite{ARXIV07064119} agrees to some extent
with the expression proposed in Refs.~\cite{DAP507p404,PRD35p1633}.
Like in our Ref.~\cite{ARXIV07064119}, the authors of Ref.~\cite{PRD35p1633}
based their ansatz on the OZI rule \cite{OZI} and the spectator picture,
finding that the production amplitude can be written as a linear
combination of the elastic and inelastic two-body scattering amplitudes,
with coefficients that do not carry any singularities, but are rather supposed
to depend smoothly on the total CM energy of the system.

However, Ref.~\cite{PRD35p1633} concluded from the unitarity relation
\begin{equation}
\Imag{A}\; =\; T^{\ast}\, A
\label{ImaTstera}
\end{equation}
that the production amplitude must be given by a \em real \em \/linear
combination of the elements of the transition matrix. A similar conclusion,
based on a $K$-matrix parametrisation, can be found in Ref.~\cite{EPJC30p503}.
In contrast, we arrive at a different conclusion, namely that
the coefficients must be {\em complex\em}, in agreement with experiment
\cite{PLB585p200,IJMPA20p482,PLB653p1} as well as with the work of the
Ishidas \cite{PTP99p1031,AIPCP688p18}.

Relation~(\ref{ImaTstera}), which can also be found in
Ref.~\cite{NPA679p671}, basically stems from the operator relations
$AV=(1+TG)V=V+TGV=T$, the symmetry of $T$, the realness of $V$,
and the unitarity of $1+2iT$,
which gives $\Imag{A}V=\Imag{AV}=\Imag{T}=T^{\ast}T=T^{\ast}AV$.
This leads, for non-singular potentials $V$,
to relation~(\ref{ImaTstera}).

Now we shall show that $A$ and $T$ satisfy the unitarity
relation~(\ref{ImaTstera}) \em despite \em \/the complexness
of the coefficients in the expansion of $A$ in terms of $T$.
Thereto, we are going to strip the expressions of Ref.~\cite{ARXIV07064119}
of all details which might obscure the simplicity of our arguments.
Hence,
let $Z_{k}(E)$ ($k=1$, 2, $\dots$, $n$)
represent a vector of complex non-singular expressions\footnote{
In Appendix~\ref{defZ} we give the precise relation
between the expressions used in Ref.~\cite{ARXIV07064119} and $Z_{k}$.},
being smooth functions of the energy $E$,
where $n$ represents the number of coupled scattering channels
under consideration,
and let the relation between $A$ and $T$ be given
by\footnote{Note that for $\Real{Z_{k}}=0$ one obtains an expansion
with real coefficients, as in Refs.~\cite{DAP507p404,PRD35p1633}.}
\begin{equation}
A_{k}\; =\;\Real{Z_{k}}\; +\; i\,\sum_{\ell}\, Z_{\ell}\, T_{k\ell}
\;\;\; .
\label{Production}
\end{equation}

We then find for the imaginary part of the production amplitude
\begin{eqnarray}
\Imag{A_{k}} & = &
\fnd{1}{2i}\,\left( A_{k}\, -\, A_{k}^{\ast}\right)
\; =\;
\sum_{\ell}\,
\left\{\Real{Z_{\ell}}\,\Real{T_{k\ell}}\; -\;
\Imag{Z_{\ell}}\,\Imag{T_{k\ell}}\right\}
\;\;\; .
\label{Step2}
\end{eqnarray}
Next, we substitute on the right-hand side of Eq.~(\ref{Step2})
the identity
$\Real{T_{k\ell}}=T_{k\ell}^{\ast}+i\Imag{T_{k\ell}}$,
and furthermore
insert the unitarity condition for $T$, {\it i.e.},
$\Imag{T_{k\ell}}=\dissum{\ell '}{}T_{\ell '\ell}T_{k\ell '}^{\ast}$,
so as to obtain
\begin{eqnarray}
\Imag{A_{k}} & = &
\sum_{\ell}\,
\left\{\Real{Z_{\ell}}\,\left( T_{k\ell}^{\ast}\, +\, i\Imag{T_{k\ell}}\right)
\; -\;\Imag{Z_{\ell}}\,\Imag{T_{k\ell}}\right\}
\nonumber \\ [10pt] & = &
\sum_{\ell}\,
\left\{\Real{Z_{\ell}}\, T_{k\ell}^{\ast}\; +\;
i\, Z_{\ell}\,\sum_{\ell '}\, T_{\ell '\ell}T_{k\ell '}^{\ast}\right\}
\;\;\; .
\label{Step3}
\end{eqnarray}
Finally, we interchange $\ell$ and $\ell '$
in the second term on the right-hand side of Eq.~(\ref{Step3}),
leaving us, also using Eq.~(\ref{Production}), with
\begin{equation}
\Imag{A_{k}}\; =\;
\sum_{\ell}\, T_{k\ell}^{\ast}
\left\{\Real{Z_{\ell}}\; +\;
i\,\sum_{\ell '}\, Z_{\ell '}\, T_{\ell\ell '}\right\}\; =\;
\sum_{\ell}\, T_{k\ell}^{\ast}\, A_{\ell}
\;\;\; .
\label{Step4}
\end{equation}
This completes the proof that $A$, as defined in Eq.~(\ref{Production}),
satisfies the unitarity condition (\ref{ImaTstera}).
Consequently, relation~(\ref{ImaTstera}) does not impose
a realness condition on the coefficients in Eq.~(\ref{Production}).

The first term on the right-hand side of relation (\ref{Production})
was not considered in
Refs.~\cite{DAP507p404,PRD35p1633,EPJC30p503}.
However, in the works of Graves-Morris \cite{NCA50p681}
and Aitchison \& collaborators \cite{NPB97p227,JPG3p1503,PLB84p349},
the possible existence of an additional real contribution was anticipated.
In Refs.~\cite{ARXIV07064119,JPG34p1789}, this follows straightforwardly from
the reasonable assumption that a produced meson pair originates from an
initial $q\bar{q}$ pair. As a consequence, the observed phenomenological
necessity \cite{PLB653p1} to employ complex coefficients in experimental
analyses of production processes does \em not \em \/allow by itself
to draw conclusions on the inevitability of including rescattering diagrams
with the spectator particle in theoretical approaches.

\section*{Acknowledgements}

We wish to thank I.~J.~R. Aitchison, D.~V.~Bugg, C.~Hanhart and
M.~R.~Pennington for useful discussions.
This work was supported in part by the {\it Funda\c{c}\~{a}o para a
Ci\^{e}ncia e a Tecnologia} \/of the {\it Minist\'{e}rio da Ci\^{e}ncia,
Tecnologia e Ensino Superior} \/of Portugal, under contract
PDCT/ FP/\-63907/\-2005.

\appendix

\section{Precise definition of \bm{Z_{k}(E)}}
\label{defZ}

In Ref.~\cite{ARXIV07064119} we discussed the partial-wave expansion
of the amplitudes for two-meson production --- together with a spectator
particle --- and scattering, assuming $q\bar{q}$ pair creation.
Hence, the coefficients bear reference to the partial wave $\ell$
and the flavor content  $\alpha$ of the quark pair.
We obtained \cite{ARXIV07064119} the following relation between production and
scattering partial-wave amplitudes:
\begin{equation}
A^{(\ell )}_{\alpha i}\; =\;
g_{\alpha i}\,
j_{\ell}\left( p_{i}r_{0}\right)\,
\sqrt{\mu_{i}\, p_{i}\,}
\, +\,
i\,\sum_{\nu}\,
g_{\alpha\nu}\,
\sqrt{\mu_{\nu}\, p_{\nu}\,}\,
h^{(1)}_{\ell}\left( p_{\nu}r_{0}\right)\,
T^{(\ell )}_{i\nu}
\;\;\; .
\label{PAmpdef}
\end{equation}
Accordingly, we must define
\begin{equation}
Z^{(\ell )}_{\alpha k}(E)\; =\;
g_{\alpha k}\,
h^{(1)}_{\ell}\left( p_{k}r_{0}\right)\,
\sqrt{\mu_{k}\, p_{k}\,}
\;\;\; .
\label{Zdef}
\end{equation}
In the latter equations, $j_\ell$ and $h^{(1)}_{\ell}$ stand for the
spherical Bessel function and  Hankel function of the first kind, respectively.
These are smooth functions of the total CM energy, just like
$\mu_{k}$ and $p_{k}$, which are the reduced mass and relative linear
momentum of the two-meson system in the $k$-th channel, respectively.
The constants $g_{\alpha k}$ stand for
the intensities of the $q\bar{q}\to MM$ couplings.
A distance scale $\sim\!0.6$ fm (for light quarks) is represented by $r_{0}$.
In the text we have stripped $Z$ of a reference
to $\ell$ and $\alpha$.

Note, moreover, as can be easily seen
from expressions (\ref{Production}) and (\ref{PAmpdef}),
that the singularity structures of the production and scattering
amplitudes are identical, since $\Real{Z_{k}}$, which is proportional to
the spherical Bessel function in Eq.~(\ref{PAmpdef}),
is a smooth function of the total invariant mass.

\newcommand{\pubprt}[4]{{#1 {\bf #2}, #3 (#4)}}
\newcommand{\ertbid}[4]{[Erratum-ibid.~{#1 {\bf #2}, #3 (#4)}]}
\def\AIPCP{AIP Conf.\ Proc.}
\def\DAP{Annalen Phys.}
\def\EPJC{Eur.\ Phys.\ J.\ C}
\def\IJMPA{Int.\ J.\ Mod.\ Phys.\ A}
\def\JPG{J.\ Phys.\ G}
\def\NCA{Nuovo Cim.\ A}
\def\NPA{Nucl.\ Phys.\ A}
\def\NPB{Nucl.\ Phys.\ B}
\def\PLB{Phys.\ Lett.\ B}
\def\PRD{Phys.\ Rev.\ D}
\def\PTP{Prog.\ Theor.\ Phys.}

\end{document}